# Self-induced passive nonreciprocal transmission by nonlinear bifacial dielectric metasurfaces

Boyuan Jin and Christos Argyropoulos*

Department of Electrical and Computer Engineering, University of Nebraska-Lincoln,

Lincoln, NE, 68588, USA

*christos.argyropoulos@unl.edu

**Abstract**

*The breaking of Lorentz reciprocity law is a non-trivial task, since it usually requires bulky magnets or complicated time-modulation dynamic techniques to be accomplished. In this work, we present a simple and compact design of a nonlinear bifacial dielectric metasurface to achieve strong self-induced passive nonreciprocal transmission without the use of external biases. The proposed design is ideal for free space optics applications, can operate under both incident polarizations, and require very low input excitation power to reach the nonreciprocal regime. It is composed of two passive silicon-based metasurfaces exhibiting Fano and Lorentzian resonances embedded in an ultrathin glass substrate. Highly asymmetric field enhancement is achieved with the proposed design that leads to strong nonreciprocity at low excitation intensities due to the large Kerr nonlinearity of silicon. Moreover, cascade designs are presented to further improve the insertion loss, broaden the nonreciprocal intensity range, and increase the isolation ratio by enhancing the transmission contrast. Finally, it is demonstrated that the proposed nonlinear metasurface is robust to fabrication imperfections and can achieve large isolation for a relative broad input power range even in the case of two*

*incident waves impinging at the same time from both directions. The current work is expected to lead to new compact nonreciprocal nanophotonic devices, such as all-optical diodes, isolators, circulators, and ultrathin protective layers for sensitive optical components.*

## I. INTRODUCTION

Nonreciprocal transmission is the fundamental operation mechanism behind all-optical isolators, circulators, and diodes [1-10]. It can control the direction of wave propagation and enables critical optical functionalities, such as laser source protection [11,12], topological beam routing and splitting [13-15], phase shifting [16,17], sensing [18,19], quantum computing [20,21]. In addition, the effect of nonreciprocity protects the transmission fidelity against possible signal instabilities, thus becoming an important functionality in the emerging field of quantum optical communications, ensuring coherent information processing [22]. Generally, time-reversal symmetric optical systems constructed by linear time-invariant materials are reciprocal, as long as an external bias is not applied [8]. Interestingly, several ways exist to break reciprocity by applying magnetic material bias [23-25], dynamic space and time modulation [26-28], and optical nonlinearity [29-36]. Although the use of magnetic materials is the most widely used technique, magnets are bulky, lossy, and expensive, making this approach difficult to be implemented on chip integrated nanophotonic circuits. In addition, the use of time modulation can be challenging to be applied in optical frequencies due to the weak response and increased complexity of electro-optical modulators [37]. On the other hand, nonreciprocity by optical nonlinear effects is a more appealing technique due to the absence of any kind of external bias. The nonlinear nonreciprocal system is self-biased by the signal itself traveling through the device. Moreover, the optical nonlinear systems do not require any active (gain) materials to boost their nonreciprocal response [34], meaning that they are completely passive and do not suffer from instabilities or other quantum noise problems.

Silicon is an ideal nonlinear material with strong third order Kerr nonlinearity [38] that has been widely used in the design of dielectric metasurfaces [39]. The fabrication of these systems is compatible with the well-established complementary metal–oxide–semiconductor (CMOS) technology used to build compact and integrated nanodevices. To achieve strong nonreciprocal transmission, the nonlinear system must be highly asymmetric, meaning that the field distribution when excited from opposite directions is different. Since the Kerr effect is proportional to the optical intensity [38], the difference in the induced nonlinear permittivity from each incident direction can lead to nonreciprocal transmittance. Fano resonators are commonly used to achieve nonlinear nonreciprocity [40-42]. They usually exhibit a structural asymmetry and a steep change in the transmission spectra, and thus can become very sensitive to the input intensity from different incident directions via nonlinear effects. However, it has been proven that a single nonlinear Fano resonator has a fundamental bound between the insertion loss and the nonreciprocal intensity contrast range [9,43,44], which limits its performance. Furthermore, the majority of the previously proposed nonlinearity-based nonreciprocal photonic devices are relative thick and usually operate in waveguide configurations. As a result, they cannot be used as free-standing structures that are ideal for free-space optics applications. Moreover, they usually exhibit poor nonreciprocal transmission contract and/or require impractical very high input intensity values to excite the aforementioned nonreciprocal effect. Finally, they cannot be used when excited by both directions in the same time due to the 'dynamic' reciprocity problem [32].

In this work, we propose a simple passive nonreciprocal device of a bifacial dielectric metasurface made of silicon spheres embedded in a glass substrate. The two metasurfaces have different geometries and function as coupled resonators, one of which has a Fano and the other a Lorentzian resonant response. The total system has a high-Q resonant response, which is ideal to achieve nonlinearity-triggered nonreciprocal response. The design is found to be robust against fabrication imperfections and disorder. Note that similar nonreciprocal performance can be obtained if the spherical silicon resonators are replaced by cylinders or other comparable geometries. Large nonreciprocal transmission at the technologically interesting telecommunication near-infrared (IR) wavelength range is realized when the input radiation is launched from opposite directions. Silicon is used in the presented design due to its large refractive index and strong nonlinear Kerr coefficient. The boosted field enhancement at the resonance of the proposed structure leads to very low required input intensity values to obtain significant nonreciprocal transmission on the order of few $MW/cm^2$. These low values will not cause saturation or other damage-leading detrimental effects, since the structure is all-dielectric with minimum optical loss and, as a result, extremely low induced ohmic heating.

To further improve the nonreciprocal performance of the proposed design, including insertion loss, nonreciprocal intensity range, isolation ratio, and flatness of nonreciprocal transmittance over a broader input intensity range, we increase the geometrical asymmetry of the existing structure by adding more metasurfaces in the presented composite nanosystem. Thus, the proposed two-layer bifacial metasurface is extended to four metasurface layers, always embedded inside a glass substrate. Two kinds of four-layer composite metasurfaces are investigated: a) cascaded pair of two coupled Fano-Lorentz bifacial metasurfaces, b) a pair of Lorentz metasurfaces followed by another pair of Fano metasurfaces. It is demonstrated that several limitations of nonreciprocal systems based on single nonlinear resonators can be overcome by these more elongated, but still ultrathin and compact, configurations [9,43,44]. Finally, due to the 'dynamic' reciprocity inherent problem of nonlinear nonreciprocal systems [32], the proposed metasurfaces are capable to work only for pulsed illumination. However, it is demonstrated that large nonreciprocal transmission still occurs when two input waves are simultaneously illuminated from opposite directions, as long as their input intensities do not exceed a moderate value, hence, relaxing the 'dynamic' reciprocity problem [32]. The presented work will lead to the design of several new compact unidirectional nanophotonic components, such as all-optical diodes, isolators, circulators, and ultrathin protective layers to decrease the damage of sensitive optical equipment from 'stray' laser signals.

## II. GEOMETRY AND LINEAR TRANSMISSION

The geometry of the proposed bifacial metasurface is shown in Fig. 1. It is composed of two opposite placed dielectric metasurfaces separated by a thin glass substrate with thickness $d = 2.2$ μm. The metasurface is made of silicon spheres embedded in glass and operates at near-IR wavelengths. The material losses are very low at this frequency range and the linear permittivities of silicon and glass are equal to $\varepsilon_{L,Si} = 12.25$ and $\varepsilon_{L,glass} = 2.1$, respectively.

The bottom metasurface is made of periodically distributed silicon spheres with radii $R_3 = 290$ nm. Due to the uniform periodic formation of silicon spheres, the bottom metasurface functions as a typical Lorentz resonator characterized by a magnetic resonance [45]. The top metasurface is composed of a bi-periodic silicon nanoparticle array interlaced by spheres with radii $R_1 = 210$ nm and $R_2 = 205$ nm. The small difference in the radii will break the in-plane symmetry, resulting to interference between different resonant modes [46] or distortion of the symmetry-protected bound states in the continuum [47], thus generating a sharp Fano resonance [48] in the transmission spectrum. However, this resonance can still prevail even for larger radii differences, as it is shown later in section IV, making the proposed design robust to fabrication imperfections. The top and bottom metasurfaces have same periodicity $a = 800$ nm along the x- and y-directions, while the nanoparticles with radii $R_1$ and $R_3$ are aligned in the z-axis as shown in the inset of Fig. 1(b), where a close-up of the proposed bifacial dielectric metasurface unit cell is presented. The forward and backward directions of the input waves impinging from the top or bottom side of the composite metasurface, respectively, are also shown in Fig. 1(b). We assume the excitation to be a plane wave traveling along the z-axis and impinging at normal incident angle on the bifacial metasurface. The incident wave can be either x- or y-polarized, leading to similar results due to the symmetric metasurface profile. During our three dimensional (3D) simulations, we modeled one unit cell of the proposed bilayer metamaterial and use periodic boundary conditions for all the lateral boundaries [49]. More details about the linear simulations can be found in [50].

The computed linear transmittance spectra of the top Fano and bottom Lorentz metasurfaces,

when the glass substrate has a thickness of $d/2$, are shown in Fig. 2. The ultrasharp Fano resonant response makes this metasurface completely transparent at $\lambda = 1524$ nm and opaque at the nearby wavelength of $\lambda = 1529$ nm. The direction and amplitude of the electric field distribution at the transmission dip ($\lambda = 1529$ nm) is shown in the inset of Fig. 2. The apparent electric field circulation is a direct signature of a magnetic dipole resonance. Interestingly, the circulation is inverse at the left and right adjacent satellite nanospheres, indicating destructive interference, which causes zero far-field transmittance due to the resulted Fano resonance [45-48]. The field enhancement is defined in this case as $|E|/E_0$, where $E$ is the local electric field and $E_0$ is the electric field amplitude of the incident wave. This Fano dielectric metasurface has a strong field enhancement. On the other hand, the bottom Lorentz metasurface has a much broader bandwidth compared to the top Fano metasurface with a peak transmittance value $T = 0.99$ at $\lambda = 1524$ nm. The field distribution of this metasurface is shown in [50] for different wavelengths. Finally, the overall linear transmittance of the proposed bifacial metasurface system, embedded in a glass substrate with thickness $d$, is depicted by the solid red line in Fig. 2. The overall transmittance cannot be derived just by multiplying the transmittance spectra of the Lorentz and Fano metasurfaces, but has a more complicated and complex shape. This is due to mutual coupling between the top and bottom metasurfaces, as well as the impact of the glass substrate thickness. The resulted transmittance spectrum of the proposed bifacial metasurface has an ideal ultrasharp shape, where the transmittance decreases rapidly from one to zero within an extremely narrow bandwidth. Such an abrupt change in transmission, combined with the enhanced and asymmetric field distributions at the top and bottom metasurfaces (computed and shown in Fig. 2 and [50]) will lead to the presented strong nonlinearity-based self-induced nonreciprocal transmission.

**III. SELF-INDUCED NONRECIPROCAL TRANSMISSION**

As the input intensity $I_0$ is increased, the Kerr nonlinear effect is expected to be triggered and alter the material properties. More precisely, it will introduce a nonlinear polarizability term given by $P_{NL} = \varepsilon_0 \chi^{(3)} |E|^2 E$, where $\varepsilon_0$ is the permittivity of free space and $\chi^{(3)}$ is the third-order nonlinear susceptibility of the material [38]. As a result, the permittivity of the material will be modulated by the local optical field intensity as $\varepsilon = \varepsilon_L + \chi^{(3)} |E|^2$, where $\varepsilon_L$ is the linear permittivity. Silicon has strong nonlinearity at near-IR range with $\chi^{(3)}_{Si} = 2.8 \times 10^{-18}$ m$^2$/V$^2$, which is four orders of magnitude larger compared to glass [38]. Moreover, the electric field is mainly confined within the silicon nanospheres, hence, we can safely neglect the Kerr nonlinear process in the glass substrate. More details about the nonlinear simulations can be found in [50].

The transmittance from both sides becomes dependent on the input intensity due to the introduction of the Kerr effect, as shown in Fig. 3, where the wavelength is fixed to $\lambda_0 = 1530$ nm. Generally, the resonance frequency would be redshifted due to the Kerr effect. The proposed bifacial metasurface is almost opaque from both incident direction illuminations when the input intensity is low, similar to the linear spectrum in Fig. 2. With the increase of the input intensity, the forward and backward transmittances will abruptly jump to a much higher level but for substantially different intensity thresholds. The threshold of the forward incident case is $I_{thF} = 1.6$ MW/cm$^2$, which is lower compared to the backward incident direction threshold $I_{thB} = 3$ MW/cm². Thus, when the input intensity is in the intensity window $I_{thF} \leq I_0 < I_{thB}$, the forward transmittance is high while the backward transmittance is near zero, exhibiting strong nonreciprocity. In addition, the fluctuation in the forward transmission is approximately $\Delta T_F = \left[\max(T_F) - \min(T_F)\right]_{I_{thF} < I_0 < I_{thB}} = 0.21$ in this nonreciprocal intensity window. We define the nonreciprocity intensity range ($NRIR = I_{thB} / I_{thF}$) to be equal to the intensity ratio where the system exhibits large nonreciprocity [44]. In the current case, the NRIR is equal to $NRIR = I_{thB} / I_{thF} = 1.9$. The transmission contrast ratio $T_F / T_B$ reaches its maximum value at $I_0 = 2.99$ MW/cm², where $T_F = 0.9$ and $T_B = 2.9 \times 10^{-3}$, ideal values for optical diode applications. This performance is substantially improved compared to all relevant devices in the literature exhibiting self-induced nonreciprocal transmission [40-42].

The nonreciprocal threshold intensities are different in the case of forward and backward illumination due to the asymmetric electric field distribution when the structure is illuminated from opposite directions. The electric field enhancement distribution along the x-y plane and across the nanosphere centers of the Fano metasurface, under a fixed input intensity $I_0 = 2.99$ MW/cm², is shown in the inset of Fig. 3. When the input is along the forward direction, the electric field enhancement in the smaller satellite nanospheres ($R_2 = 205$ nm) is stronger compared to the central nanosphere ($R_1 = 210$ nm). On the contrary, the maximum field enhancement is achieved in the central nanosphere ($R_1 = 210$ nm) in the case of backward wave illumination. In addition, the maximum field enhancement $|E|/E_0$ obtained along the lattice of the Fano metasurface is substantially increased in the forward illumination scenario, an even more important feature in order to achieve the presented nonreciprocal response. The input direction has less effect on the field distribution of the Lorentz metasurface (results shown in [50]). Hence, the Fano metasurface plays a pivotal role in the presented strong self-induced nonreciprocal response.

As discussed before, the source of the nonreciprocal transmission is the geometrical asymmetry combined with the strong nonlinearity of the proposed structure. To further increase the asymmetry in the structure's geometry, and, as a result, improve the nonreciprocal transmission contrast, two bifacial metasurfaces are used in a cascade configuration with a unit cell shown in Fig. 4(a). The dimensions of each composite metasurface are the same with those used before in Fig. 2. The system is still embedded in a glass substrate for practical reasons. The distance between the two cascade metasurfaces is chosen to be $D = 2.35$ μm, which is close to the thickness of the glass substrate of each metasurface ($d = 2.2$ μm). The forward $T_F$ and backward $T_B$ transmittances with respect to increased input intensity values are depicted in Fig. 4(b), plotted at the same input wavelength $\lambda_0 = 1530$ nm compared to Fig. 3 single metasurface design. The insertion loss is slightly decreased and maximum transmission contrast is achieved by using lower input intensities, since the maximum forward transmittance is slightly increased to $T_F = 0.91$ for $I_0 = 2.3$ MW/cm$^2$, compared to $T_F = 0.9$ in Fig. 3 at $I_0 = 2.99$ MW/cm$^2$. In addition, the flatness of the forward transmittance in the nonreciprocal input intensity range is substantially improved and becomes equal to $\Delta T_F = 0.11$. In this cascade configuration, the backward transmission is strongly suppressed compared to the single bifacial metasurface case, and this effect leads to an increased isolation ratio. In particular, the backward transmittance reaches its minimum value $T_B = 2.7 \times 10^{-4}$ at $I_0 = 2.3$ MW/cm², which is ten times lower compared to the single bifacial metasurface design with results shown in Fig. 3. The forward and backward input intensity thresholds are $I_{thF} = 1.64$ MW/cm² and $I_{thB} = 2.31$ MW/cm² in the cascade scenario, leading to $NRIR = 1.41$. The nonreciprocal range (quantified by NRIR) is slightly deteriorated compared to the single bifacial metasurface case. However, the isolation ratio and insertion loss are significantly improved with the cascade configuration.

An alternative bifacial multilayer metasurface to further increase the geometric asymmetry is shown in Fig. 5(a), where two pairs of Fano and Lorentz metasurfaces are subsequently stacked, again embedded in a glass substrate. In order to move its resonance to the same wavelength ($\lambda = 1529$ nm) used in Fig. 2, the interlayer distances are selected as $d_F = 1.8$ μm between the Fano metasurfaces, $D_p = 2.4$ μm between the Fano and Lorentz metasurfaces, and $d_L = 1.95$ μm between the Lorentz metasurfaces, respectively. All other dimensions are similar to Fig. 2 design. The computed forward $T_F$ and backward $T_B$ transmittances as functions of the increased input intensity values is shown in Fig. 5(b), while the input wavelength is equal to $\lambda_0 = 1530$ nm. The threshold intensity of the backward direction propagation is lower than that of the forward direction, i.e., $I_{thB} < I_{thF}$, where $I_{thB} = 1.3$ MW/cm² and $I_{thF} = 1.85$ MW/cm², which consists an inverse response compared to the previous scenario demonstrated in Fig. 3. Therefore, we swap the numerator and denominator in the definition of NRIR becoming: $NRIR = I_{thF} / I_{thB} = 1.42$. This NRIR value is slightly less than that of the single bifacial metasurface, but larger compared to the 4-layer composite metasurface shown before in Fig. 4(a). In this nonreciprocal window, the backward wave will be almost fully transmitted but the forward wave will be very low. The insertion loss is kept low in the NRIR with an average backward transmittance equal to $T_B = 0.99$. In addition, $T_B$ is extremely flat along the NRIR with a very small backward transmittance fluctuation $\Delta T_B = [\max(T_B) - \min(T_B)]_{I_{thB} < I_0 < I_{thF}} = 0.02$. When the input intensity is below the threshold value $I_{thF}$, the forward transmittance is kept at very low levels with a minimum value of $T_F = 0.03$ at $I_0 = 1.84$ MW/cm$^2$. This response leads to an ultrahigh isolation ratio in the entire NRIR with even lower input intensity values compared to the single bifacial metasurface design.

Normally, improving the insertion loss and the flatness of transmittance would result in a decreased NRIR. For instance, a tradeoff relation exists between the NRIR and the maximum transmittance in the nonreciprocal region for any system that contains a single nonlinear Fano resonator, which is given by $T_{\max} \leq T_{\lim} = 4NRIR/(NRIR+1)^2$ [9,43,44]. In the case of single bifacial metasurface design shown in Fig. 3, we can derive that $NRIR = 1.9$, leading to a transmittance limit $T_{\lim} = 0.9$. The maximum nonreciprocal transmittance in this case is $T_{\max} = 0.9$, exactly at this limit. In the metasurface shown in Fig. 4(a), the maximum nonreciprocal transmittance does not exceed the transmittance limit: $T_{\max} = 0.91 < T_{\lim} = 0.97$ with $NRIR = 1.41$. On the contrary, the bifacial multilayer metasurface composed of two pairs of Fano and Lorentz metasurfaces shown in Fig. 5(a) can break this limit, achieving $T_{\max} = 0.99 > T_{\lim} = 0.97$. This is due to the fact that the composite metasurface of Fig. 5(a) can significantly increase the flatness and transmittance maximum. The composite design has the best performance compared to all previously presented self-induced nonreciprocal devices [40-42] and can be employed as a compact free-standing optical isolator or diode.

In the results presented in Fig. 3, it is assumed that at any given time, only one pulse propagates along the metasurface, either from the forward or backward direction. Hence, the proposed bifacial metasurface can exhibit strong nonreciprocal transmission when illuminated by a pulsed wave in order to eliminate the 'dynamic' reciprocity problem [32]. However, it will be interesting to investigate the presented self-induced nonreciprocity effect when two waves are simultaneously launched from opposite directions and coexist inside the metasurface, as schematically shown in Fig. 6(a). The wave with intensity $I_{in1}$ and power $P_{in1} = a^2 I_{in1}$ propagates in the forward direction, while the wave with intensity $I_{in2}$ and power $P_{in2} = a^2 I_{in2}$ travels opposite, where $a$ is the structure's period. The measured total output powers from the bottom and top side of the bifacial metasurface are $P_{out1}$ and $P_{out2}$, respectively. The subscripts 1 and 2 represent the forward and backward propagation direction, respectively. We assume a constant input intensity value $I_0 = 2.3$ MW/cm² that is located at the center of the NRIR in Fig. 3, where large nonreciprocity exists under single illumination with $T_F = 0.86$ and $T_B = 0.037$. Then, we assign the forward input intensity to have this fixed value $I_{in1} = I_0$, while increasing the input intensity $I_{in2}$ of the backward input wave. In this scenario, the corresponding output power ratio at the bottom side of the bifacial metasurface, given by $\eta_1 = P_{out1} / (P_{in1} + P_{in2})$, is computed and shown by the dashed red line in Fig. 6(b). Similarly, when $I_{in2} = I_0$ and $I_{in1}$ is increasing, the corresponding output power ratio at the top side of the bifacial metasurface, given by $\eta_2 = P_{out2} / (P_{in1} + P_{in2})$, is calculated and depicted by the solid black line in Fig. 6(b). Interestingly, if the input at the output port exceeds a threshold value of ~0.17 MW/cm$^2$, 'dynamic' reciprocity comes into effect [32] and the proposed metasurface is no longer nonreciprocal. However, Fig. 6(b) clearly demonstrates that the nonreciprocal transmission can remain as long as the input signal intensity from the opposite direction is lower than that threshold, which is ~1/10 of the threshold intensity ($I_{thF}$ or $I_{thB}$) in the case of the single excitation nonreciprocal operation.

The potential experimental verification of the proposed nonreciprocal design is feasible. The bifacial all-dielectric metasurface can be built by well-established semiconductor nanofabrication methods combined with material transfer techniques [51-53]. To further reduce the fabrication complexity, the spheres can be replaced by nanocylinders or nanocones without affecting the performance of the proposed device, since they also support Mie-like resonances [39,54]. Polydimethylsiloxane (PDMS) has refractive index similar to the currently used glass and can be used as an alternative intermediate substrate material to embed the presented silicon metasurface design [52].

## IV. ROBUSTNESS AGAINST GEOMETRICAL IMPERFECTIONS

The Fano resonance shown in Fig. 2 originates from the small difference in the silicon sphere radii of the Fano metasurface. In the current design, the radii of the silicon spheres are chosen to be $R_1 = 210$ nm and $R_2 = 205$ nm, with a small radii difference of $(R_1 - R_2) = 5$ nm. During the potential fabrication and experimental verification of the proposed concept, geometrical imperfections are expected to be induced in the metasurface's structure leading to deviations in the sphere radii difference. To further study the robustness of the proposed metasurface to potential fabrication imperfections, we keep the radii average $(R_1 + R_2)/2$ constant while varying the radii difference $(R_1 - R_2)$ from 5 nm to 105 nm. The computed transmittance spectra of the proposed linear bifacial metasurfaces by using substantially different radii dimensions are shown in Fig. 7(a). All the other parameters are kept the same with the ones used in the design of Fig. 2. When the radii difference is increased, the transmission peak declines and the resonance wavelength slightly red-shifts. In addition, the resonance bandwidth becomes broader [47]. However, the linear transmission peak can be easily tuned back to unity and the resulted Fano resonance can be recovered, as shown in Fig. 7(b), even in the extremely disordered and imperfect metasurface with dimensions $R_1 = 260$ nm, $R_2 = 155$ nm, by properly adjusting the thickness of the glass substrate, which now becomes slightly larger $d = 2.3$ μm compared to the $d = 2.2$ μm value used in Fig. 2.

The computed linear transmittance spectrum in Fig. 7(b) is not as sharp as that presented in Fig. 2. However, the abrupt transitions in the calculated forward and backward nonreciprocal transmittances are still preserved when considering the nonlinear Kerr effect. The nonlinear forward $T_F$ and backward $T_B$ transmittances are plotted in Fig. 7(c) as functions of the input intensity in the case of the imperfect bifacial metasurface with dimensions $R_1 = 260$ nm, $R_2 = 155$ nm, and $d = 2.3$ μm. The wavelength is chosen to be $\lambda_0 = 1545$ nm in this case, which is very close to the nonreciprocal operation wavelength of the previous designs. Due to the highly asymmetric enhanced electric field distributions, $T_F$ and $T_B$ can abruptly transit from low to high values at significantly different input intensities as shown in Fig. 7(c), similar to the nonreciprocal response of the ideal structure shown in Fig. 3. The thresholds for $T_F$ and $T_B$ are $I_{thF}=92$ MW/cm$^2$ and $I_{thB}=44$ MW/cm$^2$, respectively, in this case, which are slightly larger values compared to the input intensities used in the ideal design of Fig. 3. Large contrast between forward and backward transmittances is obtained for any input intensities between these thresholds. The maximum contrast occurs at $I_0=44$ MW/cm$^2$, where $T_F=0.08$ and $T_B=0.82$. The main difference of the currently presented imperfect bifacial metasurface design (Fig. 7(c)) compared to the ideal design of Fig. 3 is that higher input intensities are required to trigger nonreciprocal transmission. However, the strong nonreciprocal response is still present. Hence, we can conclude that the proposed bifacial metasurface is robust against imperfections expected to be induced during the fabrication process.

To mimic a more complex geometrical disorder in the metasurface geometry, Fig. 8(a) represents the schematic of a more complicated rectangular supercell design with a period of length $2a$ and width $a$. The bottom Lorentz metasurface is kept the same to that used in Fig. 1(b), since this metasurface is easier to be fabricated and, as a result, is expected to be less affected by geometrical imperfections [39,54]. The upper Fano metasurface is composed by two silicon spheres at the center with $R_1$=210 nm, two half-spheres on the up and down edges with $R_4$=215 nm, and four quarter-spheres at the corners with different and smaller size compared to $R_4$, now equal to $R_2$=205 nm. The thickness of the glass substrate is $d$=1985 nm. It can be seen in Fig. 8(b) that this disordered and imperfect bifacial metasurface can also exhibit a Fano-like linear response, independent to the difference among the sphere radii that is possible to occur due to fabrication imperfections. The linear response is similar to the ideal metasurface response presented in Fig. 2. Interestingly, the nonlinear Kerr effect can also generate large nonreciprocal transmittance in this disordered metasurface when the wavelength is fixed to $\lambda_0 = 1524.5$ nm and the input intensity is in the range 2.4 MW/cm² $< I_0 <$ 3.5 MW/cm² with results shown in Fig. 8(c). The maximum nonreciprocal transmission contrast occurs at $I_0 = 3.5$ MW/cm², where $T_F = 0.88$ and $T_B = 0.07$. Hence, we can conclude that the proposed bifacial metasurface is also robust against potential geometrical disorder that can occur during the fabrication process, which is expected to induce different dimensions in adjacent nanoparticles.

The aforementioned different geometrical deviation effects in the sphere radii are very similar to other structural disorders that can be possibly induced during the fabrication process, such as random periodicity and imperfect alignment [47]. Hence, it is proven by Fig. 7 and 8 that the linear and nonlinear responses of the proposed bifacial metasurface are robust against geometrical disorder and imperfections. The minor drawback caused by these imperfections is that nonreciprocity can now be achieved at a slightly shifted wavelength (frequency detuning) or for marginally increased input intensity values. Therefore, the strong self-induced nonreciprocal transmission is expected to be verified by potential experimental efforts based on the concept presented in this paper.

## V. CONCLUSIONS

To conclude, an all-dielectric low-loss nonlinear bifacial metasurface is proposed that is able to achieve passive and bias-free strong nonreciprocity in transmission. The presented realistic metasurface design is made of a Lorentz and Fano resonator embedded inside a glass substrate. The highly structural asymmetry of the proposed metasurface combined with the strong field enhancement at the resonance cause strong nonreciprocal transmission due to the enhanced nonlinear Kerr effect. The required input intensities to achieve this effect are relative low, on the order of few MW/cm$^2$, an intriguing property that can lead to single- or few-photon quantum optical nonreciprocal devices [55]. To further improve the insertion loss, nonreciprocal intensity range, isolation ratio, and flatness of transmittance, two additional multilayer metasurface designs are proposed, where the tradeoff relation between insertion loss and nonreciprocal intensity range is relaxed and even outperformed. Finally, the scenario of two input waves simultaneously launched from opposite directions is studied. Large nonreciprocal transmission contrast is also dominant in this case but for a moderate range of input intensity values. The proposed ultrathin compact metasurface has a very low insertion loss, extremely flat nonreciprocal transmittance, and strong isolation over a broad nonreciprocal intensity range. It is proven that is robust to fabrication imperfections and consists an ideal design for free-space optics applications. To the best of our knowledge, none of the previously proposed self-induced nonreciprocal nonlinear devices [40-42] have realized all these desired features simultaneously in a fully passive scenario without the inclusion of active materials. Our work is expected to lead to several applications in the emerging field of compact unidirectional nanophotonic devices, such as all-optical diodes, isolators, circulators, and ultrathin protective layers to decrease the damage of sensitive optical components.

**ACKNOWLEDGEMENTS**

This work was partially supported by the NSF (Grant No. DMR‐1709612), the NSF Nebraska Materials Research Science and Engineering Center (Grant No. DMR‐1420645), Office of Naval Research Young Investigator Program (ONR‐YIP) Award (Grant No. N00014‐19‐1‐2384), and the NSF‐Nebraska‐EPSCoR (Grant No. OIA‐1557417).

**REFERENCES**

[1] R. J. Potton, Reciprocity in optics, Rep. Prog. Phys. **67**, 717 (2004).

[2] Z. F. Yu and S. H. Fan, Complete optical isolation created by indirect interband photonic transitions, Nat. Photonics **3**, 91 (2009).

[3] L. Bi, J. J. Hu, P. Jiang, D. H. Kim, G. F. Dionne, L. C. Kimerling, and C. A. Ross, On-chip optical isolation in monolithically integrated non-reciprocal optical resonators, Nat. Photonics **5**, 758 (2011).

[4] D. Jalas *et al.*, What is - and what is not - an optical isolator, Nat. Photonics **7**, 579 (2013).

[5] L. Chang, X. S. Jiang, S. Y. Hua, C. Yang, J. M. Wen, L. Jiang, G. Y. Li, G. Z. Wang, and M. Xiao, Parity-time symmetry and variable optical isolation in active-passive-coupled microresonators, Nat. Photonics **8**, 524 (2014).

[6] A. M. Mahmoud, A. R. Davoyan, and N. Engheta, All-passive nonreciprocal metastructure, Nat. Commun. **6**, 8359 (2015).

[7] D. L. Sounas and A. Alu, Non-reciprocal photonics based on time modulation, Nat. Photonics **11**, 774 (2017).

[8] C. Caloz, A. Alu, S. Tretyakov, D. Sounas, K. Achouri, and Z. L. Deck-Leger,

Electromagnetic nonreciprocity, Phys. Rev. Appl. **10**, 047001 (2018).

[9] D. L. Sounas and A. Alu, Fundamental bounds on the operation of Fano nonlinear isolators, Phys. Rev. B **97**, 115431 (2018).

[10] S. A. Mann, D. L. Sounas, and A. Alu, Nonreciprocal cavities and the time-bandwidth limit, Optica **6**, 104 (2019).

[11] S. Y. Hua, J. M. Wen, X. S. Jiang, Q. Hua, L. Jiang, and M. Xiao, Demonstration of a chip-based optical isolator with parametric amplification, Nat. Commun. **7**, 13657 (2016).

[12] F. Ruesink, M. A. Miri, A. Alu, and E. Verhagen, Nonreciprocity and magnetic-free isolation based on optomechanical interactions, Nat. Commun. **7**, 13662 (2016).

[13] S. Taravati and A. A. Kishk, Dynamic modulation yields one-way beam splitting, Phys. Rev. B **99**, 075101 (2019).

[14] S. A. H. Gangaraj and F. Monticone, Topologically-protected one-way leaky waves in nonreciprocal plasmonic structures, J. Phys. Condens. Matter **30**, 104002 (2018).

[15] S. Kruk, A. Poddubny, D. Smirnova, L. Wang, A. Slobozhanyuk, A. Shorokhov, I. Kravchenko, B. Luther-Davies, and Y. Kivshar, Nonlinear light generation in topological nanostructures, Nat. Nanotechnol. **14**, 126 (2019).

[16] C. J. Firby and A. Y. Elezzabi, High-speed nonreciprocal magnetoplasmonic waveguide phase shifter, Optica **2**, 598 (2015).

[17] Z. Shen, Y. L. Zhang, Y. Chen, C. L. Zou, Y. F. Xiao, X. B. Zou, F. W. Sun, G. C. Guo, and C. H. Dong, Experimental realization of optomechanically induced non-reciprocity, Nat. Photonics **10**, 657 (2016).

[18] L. Del Bino, J. M. Silver, S. L. Stebbings, and P. Del'Haye, Symmetry breaking of counter-

propagating light in a nonlinear resonator, Sci. Rep. **7**, 43142 (2017).

[19] H. K. Lau and A. A. Clerk, Fundamental limits and non-reciprocal approaches in non-Hermitian quantum sensing, Nat. Commun. **9**, 4320 (2018).

[20] S. Longhi, D. Gatti, and G. Della Valle, Non-Hermitian transparency and one-way transport in low-dimensional lattices by an imaginary gauge field, Phys. Rev. B **92**, 094204 (2015).

[21] Y. Tokura, M. Kawasaki, and N. Nagaosa, Emergent functions of quantum materials, Nat. Phys. **13**, 1069 (2017).

[22] P. Doyeux, S. A. H. Gangaraj, G. W. Hanson, and M. Antezza, Giant Interatomic Energy-Transport Amplification with Nonreciprocal Photonic Topological Insulators, Phys. Rev. Lett. **119**, 173901 (2017).

[23] D. N. Huang, P. Pintus, C. Zhang, P. Morton, Y. Shoji, T. Mizumoto, and J. E. Bowers, Dynamically reconfigurable integrated optical circulators, Optica **4**, 23 (2017).

[24] C. Zhang, P. Dulal, B. J. H. Stadler, and D. C. Hutchings, Monolithically-Integrated TE-mode 1D silicon-on-insulator isolators using seedlayer-free garnet, Sci. Rep. **7**, 5820 (2017).

[25] T. Mizumoto, Y. Shoji, and R. Takei, Direct wafer bonding and its application to waveguide optical isolators, Materials **5**, 985 (2012).

[26] Y. S. Yang, C. Galland, Y. Liu, K. Tan, R. Ding, Q. Li, K. Bergman, T. Baehr-Jones, and M. Hochberg, Experimental demonstration of broadband Lorentz non-reciprocity in an integrable photonic architecture based on Mach-Zehnder modulators, Opt. Express **22**, 17409 (2014).

[27] C. R. Doerr, L. Chen, and D. Vermeulen, Silicon photonics broadband modulation-based

isolator, Opt. Express **22**, 4493 (2014).

[28] K. J. Fang, J. Luo, A. Metelmann, M. H. Matheny, F. Marquardt, A. A. Clerk, and O. Painter, Generalized non-reciprocity in an optomechanical circuit via synthetic magnetism and reservoir engineering, Nat. Phys. **13**, 465 (2017).

[29] R. Philip, M. Anija, C. S. Yelleswarapu, and D. V. G. L. N. Rao, Passive all-optical diode using asymmetric nonlinear absorption, Appl. Phys. Lett. **91**, 141118 (2007).

[30] L. Fan, J. Wang, L. T. Varghese, H. Shen, B. Niu, Y. Xuan, A. M. Weiner, and M. H. Qi, An all-silicon passive optical diode, Science **335**, 447 (2012).

[31] B. Peng *et al.*, Parity-time-symmetric whispering-gallery microcavities, Nat. Phys. **10**, 394 (2014).

[32] Y. Shi, Z. F. Yu, and S. H. Fan, Limitations of nonlinear optical isolators due to dynamic reciprocity, Nat. Photonics **9**, 388 (2015).

[33] L. Del Bino, J. M. Silver, M. T. M. Woodley, S. L. Stebbings, X. Zhao, and P. Del'Haye, Microresonator isolators and circulators based on the intrinsic nonreciprocity of the Kerr effect, Optica **5**, 279 (2018).

[34] B. Y. Jin and C. Argyropoulos, Nonreciprocal transmission in nonlinear PT-symmetric metamaterials using epsilon-near-zero media doped with defects, Adv. Opt. Mater. **7**, 1901083 (2019).

[35] M. Lawrence and J. A. Dionne, Nanoscale nonreciprocity via photon-spin-polarized stimulated Raman scattering, Nat. Commun. **10**, 3297 (2019).

[36] P. Yang *et al.*, Realization of nonlinear optical nonreciprocity on a few-photon level based on atoms strongly coupled to an asymmetric cavity, Phys. Rev. Lett. **123**, 233604 (2019).

[37] K. Liu, C. R. Ye, S. Khan, and V. J. Sorger, Review and perspective on ultrafast wavelength-size electro-optic modulators, Laser Photonics Rev. **9**, 172 (2015).

[38] R. W. Boyd, *Nonlinear optics* (Academic Press, New York, 2008).

[39] A. I. Kuznetsov, A. E. Miroshnichenko, M. L. Brongersma, Y. S. Kivshar, and B. Luk'yanchuk, Optically resonant dielectric nanostructures, Science **354**, aag2472 (2016).

[40] D. L. Sounas and A. Alu, Nonreciprocity based on bonlinear resonances, IEEE Antennas Wirel. Propag. Lett. **17**, 1958 (2018).

[41] Y. Yu, Y. H. Chen, H. Hu, W. Q. Xue, K. Yvind, and J. Mork, Nonreciprocal transmission in a nonlinear photonic-crystal Fano structure with broken symmetry, Laser Photonics Rev. **9**, 241 (2015).

[42] M. Lawrence, D. R. Barton, and J. A. Dionne, Nonreciprocal flat optics with silicon metasurfaces, Nano Lett. **18**, 1104 (2018).

[43] D. L. Sounas and A. Alu, Time-Reversal symmetry bounds on the electromagnetic response of asymmetric structures, Phys. Rev. Lett. **118**, 154302 (2017).

[44] D. L. Sounas, J. Soric, and A. Alu, Broadband passive isolators based on coupled nonlinear resonances, Nat. Electron. **1**, 113 (2018).

[45] A. Garcia-Etxarri, R. Gomez-Medina, L. S. Froufe-Perez, C. Lopez, L. Chantada, F. Scheffold, J. Aizpurua, M. Nieto-Vesperinas, and J. J. Saenz, Strong magnetic response of submicron Silicon particles in the infrared, Opt. Express **19**, 4815 (2011).

[46] W. Y. Zhao, D. Q. Ju, Y. Y. Jiang, and Q. W. Zhan, Dipole and quadrupole trapped modes within bi-periodic silicon particle array realizing three-channel refractive sensing, Opt. Express **22**, 31277 (2014).

[47] K. Koshelev, S. Lepeshov, M. K. Liu, A. Bogdanov, and Y. Kivshar, Asymmetric metasurfaces with high-Q resonances governed by bound states in the continuum, Phys. Rev. Lett. **121**, 193903 (2018).

[48] C. Argyropoulos, Enhanced transmission modulation based on dielectric metasurfaces loaded with graphene, Opt. Express **23**, 23787 (2015).

[49] COMSOL MULTIPHYSICS, http://www.comsol.com/.

[50] See Supplemental Material for details about the linear and nonlinear simulations, and the computed electric field distributions.

[51] A. Shevchenko, V. Kivijarvi, P. Grahn, M. Kaivola, and K. Lindfors, Bifacial metasurface with quadrupole optical response, Phys. Rev. Appl. **4**, 024019 (2015).

[52] Y. Zhou, I. I. Kravchenko, H. Wang, J. R. Nolen, G. Gu, and J. Valentine, Multilayer noninteracting dielectric metasurfaces for multiwavelength metaoptics, Nano Lett. **18**, 7529 (2018).

[53] W. J. Liu, Q. S. Zou, C. Q. Zheng, and C. J. Jin, Metal-Assisted transfer strategy for construction of 2D and 3D nanostructures on an elastic substrate, ACS Nano **13**, 440 (2019).

[54] S. Jahani and Z. Jacob, All-dielectric metamaterials, Nat. Nanotechnol. **11**, 23 (2016).

[55] H. Choi, M. Heuck, and D. Englund, Self-similar nanocavity design with ultrasmall mode volume for single-photon nonlinearities, Phys. Rev. Lett. **118**, 223605 (2017).

**Figures**

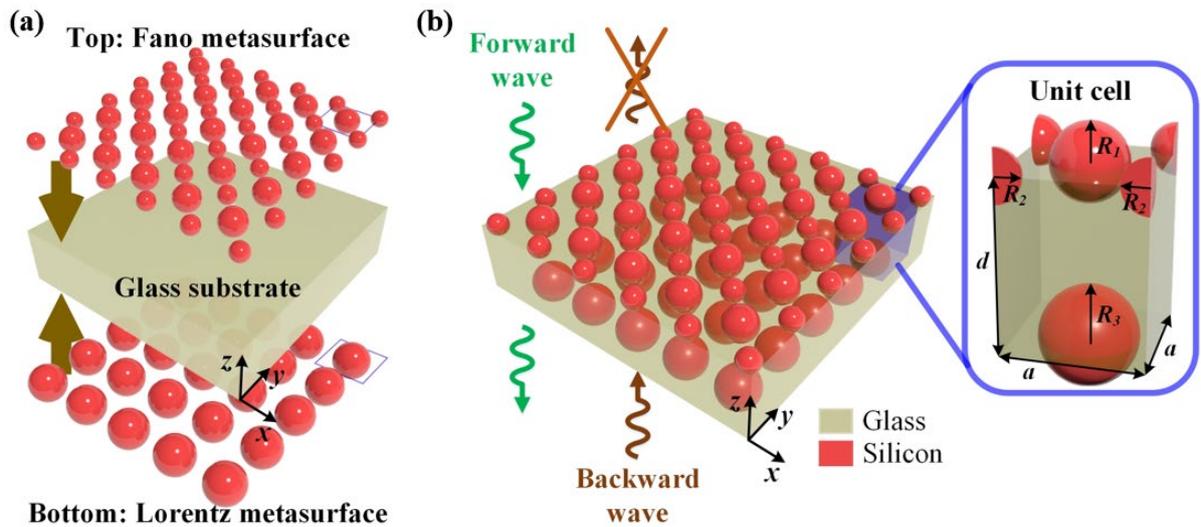

FIG. 1. (a) Proposed bifacial dielectric metasurface. (b) Schematic illustration of the nonreciprocal transmission operation. Inset: the unit cell of the composite metasurface.

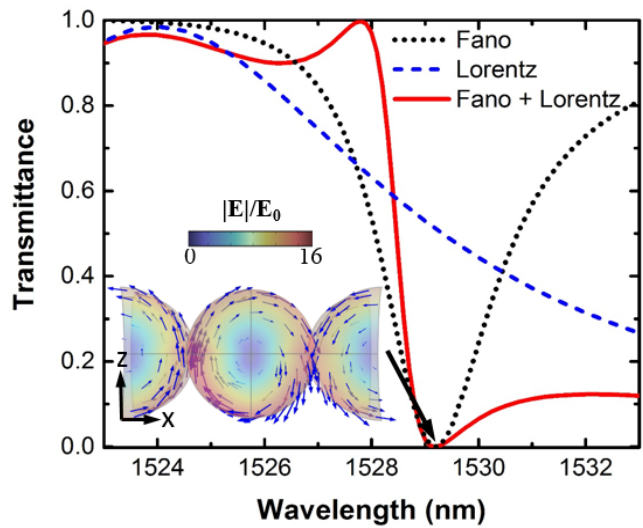

FIG. 2. Linear transmittance spectra of the top Fano metasurface (dotted black line), bottom Lorentz metasurface (dashed blue line), and the proposed bifacial metasurface (solid red line). Inset: the direction (blue arrows) and amplitude (color map) of the electric field on the silicon nanoparticles of the Fano metasurface at $\lambda = 1529$ nm, where the transmittance is equal to zero.

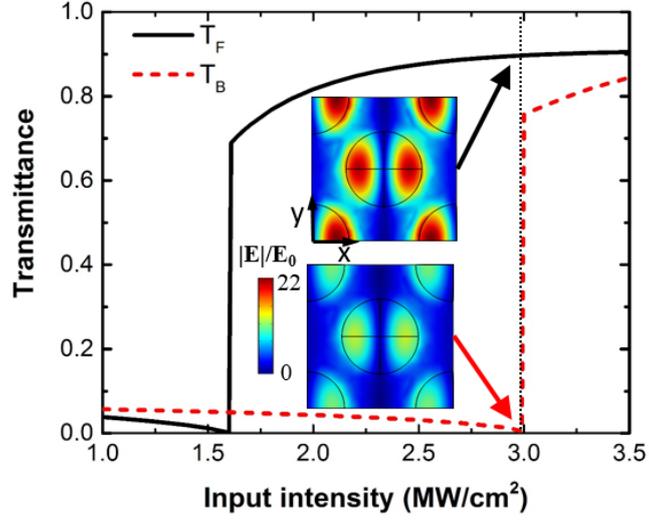

FIG. 3. Nonlinear forward $T_F$ and backward $T_B$ transmittances at $\lambda_0 = 1530$ nm as functions of the input intensity. Inset: the electric field at $I_0 = 2.99$ MW/cm$^2$ along the x-y plane, across the nanosphere centers of the Fano metasurface.

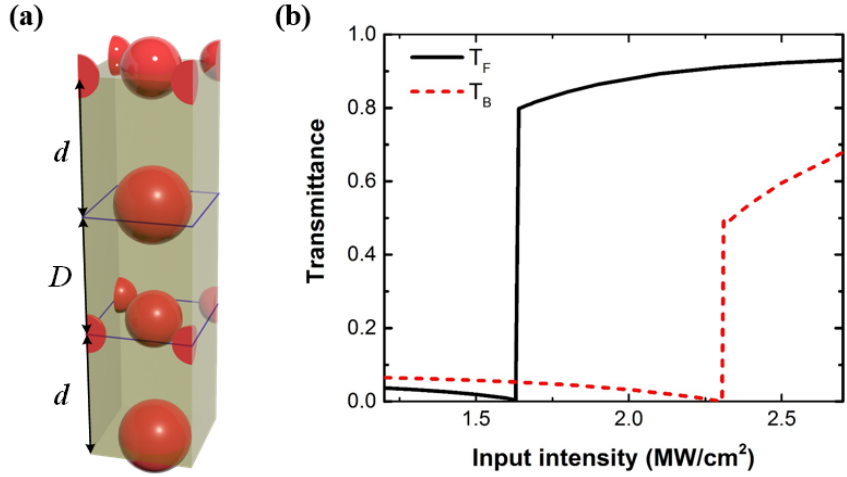

FIG. 4. (a) Cascade unit cell design made of two bifacial metasurfaces in series. (b) Nonlinear forward $T_F$ and backward $T_B$ transmittances at $\lambda_0 = 1530$ nm as functions of the input intensity.

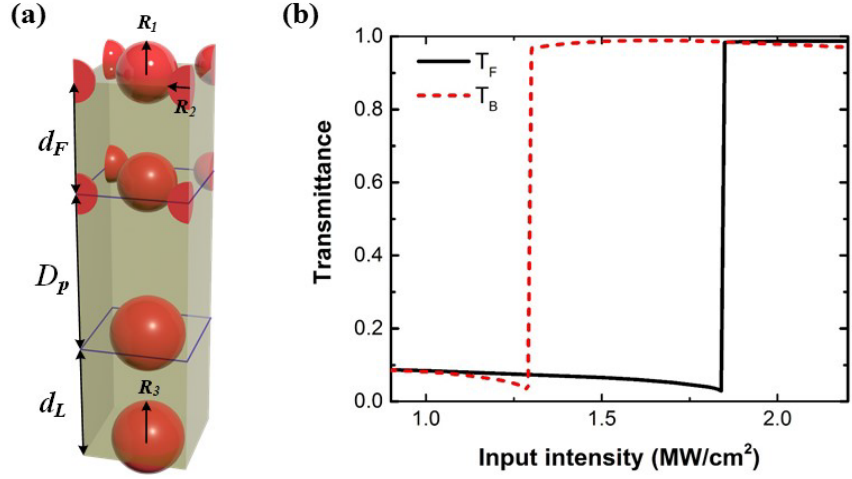

FIG. 5. (a) Alternative bifacial multilayer metasurface unit cell design, where two pairs of the Fano and Lorentz metasurfaces are subsequently stacked. (b) Nonlinear forward $T_F$ and backward $T_B$ transmittances at $\lambda_0 = 1530$ nm as functions of the input intensity.

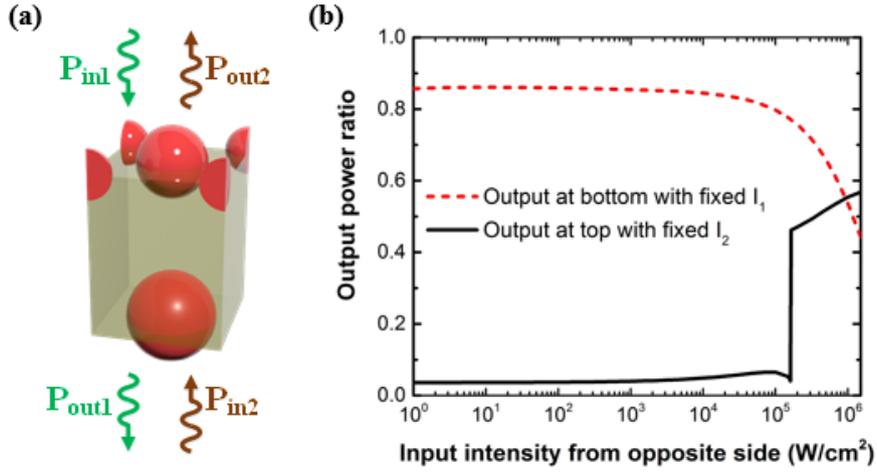

FIG. 6. (a) Single bifacial metasurface unit cell simultaneously illuminated by two input signals launched from opposite directions. (b) Nonlinear output power ratio computed at the bottom or top metasurface side when the input intensity from one direction is fixed to 2.3 MW/cm$^2$ while the intensity from the opposite direction varies.

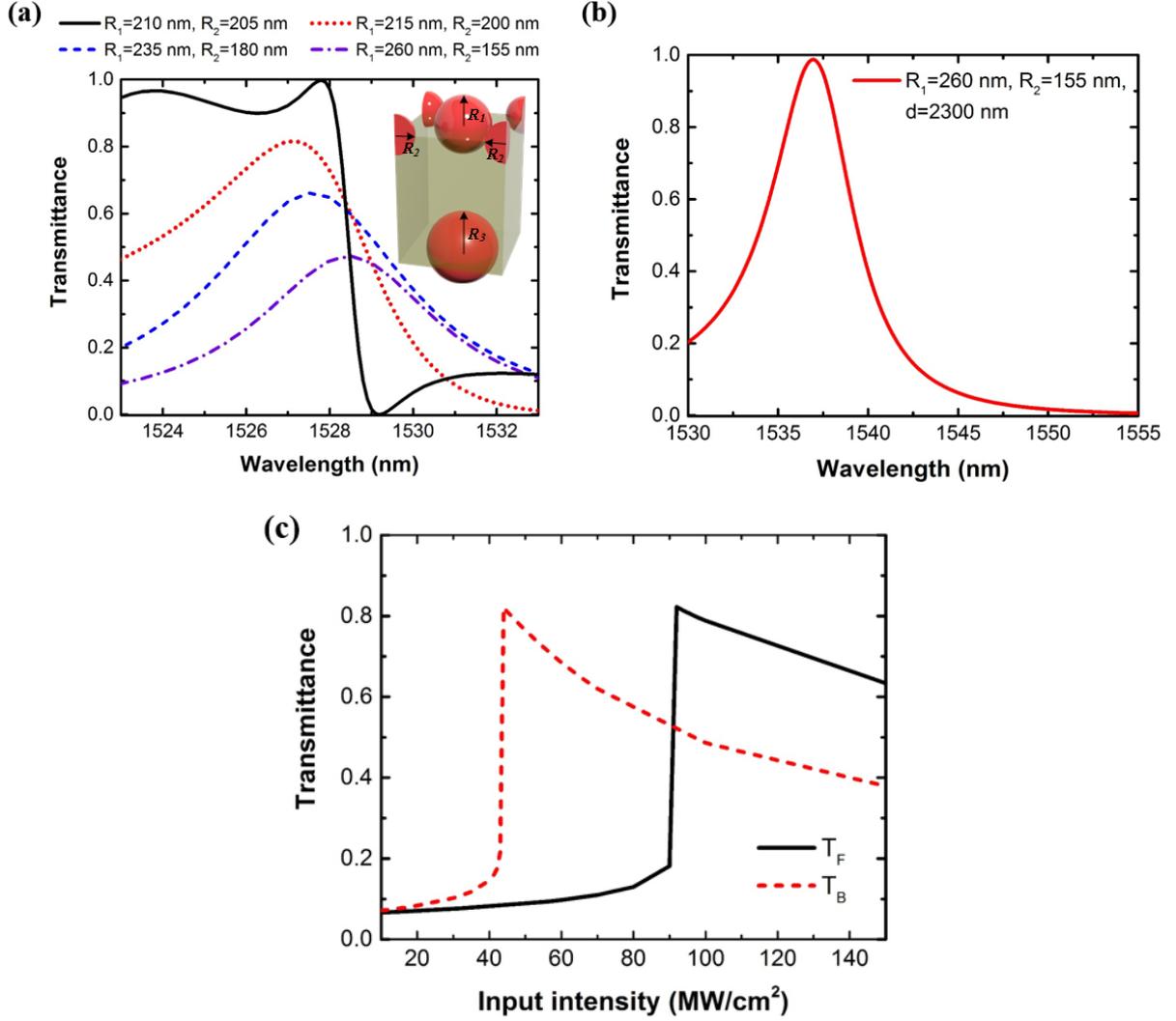

FIG. 7. (a) Linear transmittance spectra of the bifacial metasurface based on silicon spheres with varying radii in the Fano metasurface. Inset: the unit cell of the proposed bifacial metasurface. (b) Linear transmittance of the bifacial metasurface with dimensions $R_1$=260 nm, $R_2$=155 nm, and $d$=2300 nm. The unity transmittance and large contrast are recovered compared to (a) by adjusting the thickness of the glass substrate leading to similar performance compared to the ideal metasurface design (Fig. 2). (c) Nonlinear forward $T_F$ and backward $T_B$ transmittances as functions of the input intensity in the case of the imperfect bifacial metasurface geometry with dimensions given in (b). The input wavelength is chosen to be $\lambda_0 = 1545$ nm.

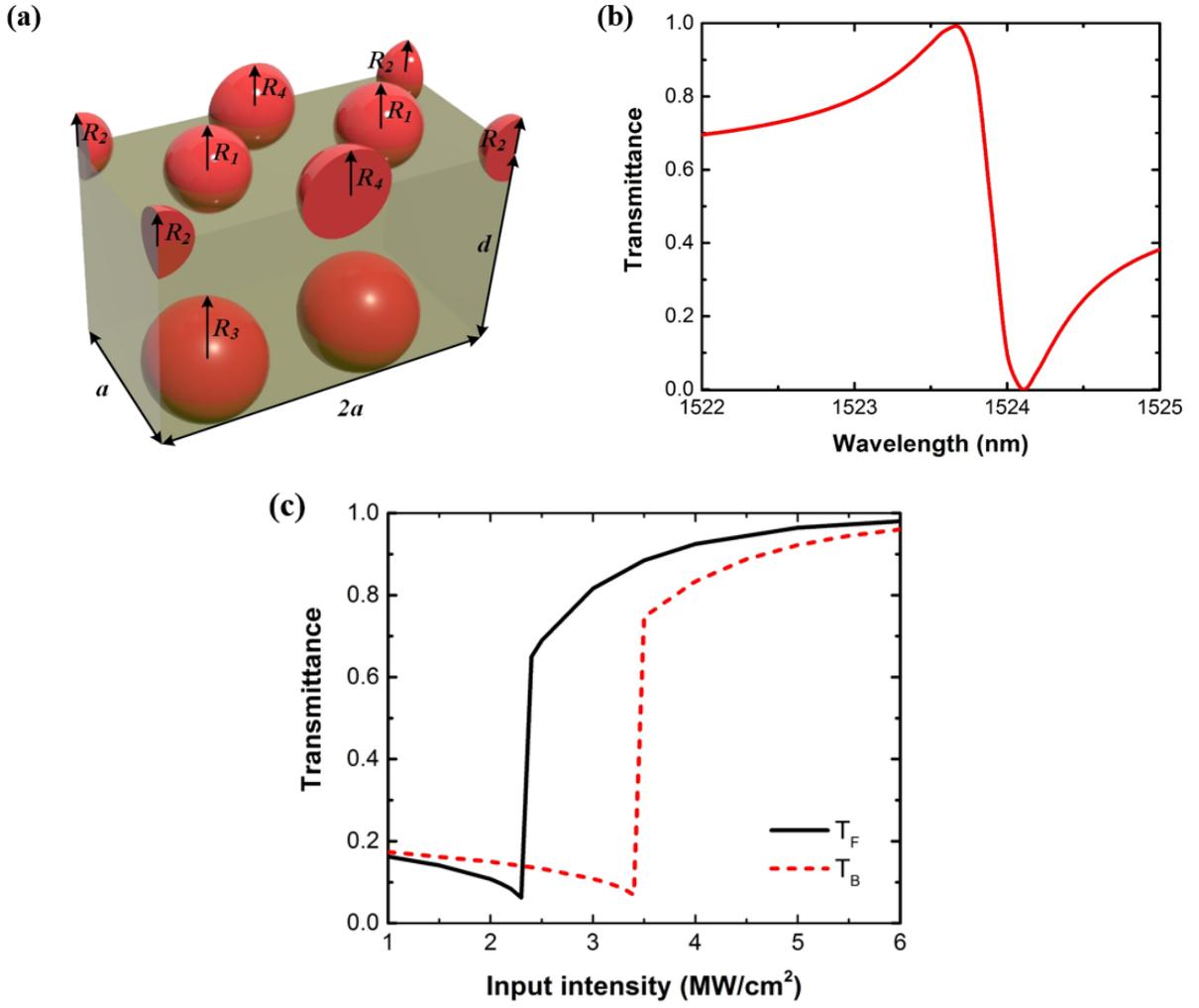

FIG. 8. (a) Schematic of a disordered metasurface supercell composed by two adjacent unit cells with $R_1$=210 nm, $R_2$=205 nm, $R_3$=290 nm, $R_4$=215 nm, and $d$=1985 nm. (b) Linear transmittance of the bifacial metasurface based on the supercell depicted in (a). The disordered metasurface also exhibits a Fano-like resonant response similar to the ideal design presented in Fig. 2. (c) Nonlinear forward $T_F$ and backward $T_B$ transmittances as function of the input intensity at $\lambda_0 = 1524.5$ nm for the disordered metasurface.

# Supplemental Material

# Self-induced passive nonreciprocal transmission by nonlinear bifacial dielectric metasurfaces

Boyuan Jin and Christos Argyropoulos*

Department of Electrical and Computer Engineering, University of Nebraska-Lincoln,

Lincoln, NE, 68588, USA

*christos.argyropoulos@unl.edu

## 1. Numerical method

In this paper, the linear and nonlinear responses of the metasurfaces are computed by COMSOL Multiphysics, a full-wave electromagnetic solver based on the finite element method (FEM). The silicon spheres are modeled as three-dimensional (3D) structures and they have a curved geometry. To reduce the computational load, the unit cell of the proposed complicated bifacial metasurfaces, shown in the inset of Fig. 1(b), is terminated by periodic boundary conditions on the lateral sides in both x- and y-directions. Port boundaries are placed several wavelengths away from the metasurfaces in the up and down sides along the z-direction to create the incident plane wave. The input waves are linearly polarized along the x-direction. Besides, it has been verified that the proposed bifacial metasurface respond similarly to y-polarized incident waves due to the symmetrical geometry.

Since we assume that the incident waves are narrowband, the linear permittivities of silicon and glass are independent of the wavelength and equal to $\varepsilon_{L,Si} = 12.25$ and $\varepsilon_{L,glass} = 2.1$, respectively. In the case of the used nonlinear Kerr effect, a nonlinear polarizability term is induced equal to: $P_{NL} = \varepsilon_0 \chi^{(3)} |E|^2 E$, where $\varepsilon_0$ is the permittivity of free space and $\chi^{(3)}$ is the third-order nonlinear susceptibility of the material. Thus, the wave equation, derived by Maxwell's equations, in the frequency domain when including the nonlinear effect becomes:

$$\nabla \times (\mu_r^{-1} \nabla \times \boldsymbol{E}) - \varepsilon_r k_0^2 \boldsymbol{E} = \mu_0 \omega^2 \boldsymbol{P}_{NL}. \qquad (1)$$

Equation (1) deviates from the standard linear wave equation formula with the addition of a non-zero nonlinear term on its right-hand side. Therefore, the master equation in COMSOL, representing the linear wave equation, must be modified to accommodate the used Kerr nonlinear effect. We have amended the COMSOL master equation and added the non-zero third-order nonlinear polarizability term by introducing a weak contribution module in our model. By solving the modified wave equation, as shown by Eq. (1), the steady state solutions of the current nonlinear problem can be attained. Resonant structures under the Kerr effect often exhibit optical bistability, suggesting two distinct steady state solutions can coexist for the same input intensity. The actual solution is determined by the continuity, which means the transmission depends on its previous value. For consistency, the input is scanned from low to high intensities throughout this work, similar to the practical case of a laser with increased input power.

The mesh is chosen to be extremely fine in the silicon nanospheres, with a minimum size of 10 nm. Silicon is the only nonlinear material in the simulated system. The extremely fine mesh can improve the convergence in the nonlinear simulations and the computational accuracy of the computed results. On the other hand, the mesh size in the glass substrate is chosen to be coarser, in order to accelerate the computation time, and becomes even coarser in the up and down free space domains as the distance from the bifacial metasurface is increased.

The reflectance $R$ and transmittance $T$ of the proposed bifacial metasurface is computed by the S-parameters by using the formulas: $R = |S_{11}|^2$ and $T = |S_{21}|^2$. Normally, COMSOL can automatically calculate the S-parameter of the system through the input (Port 1) and output (Port 2) ports. In Fig. 6, however, two input waves are simultaneously injected to the bifacial metasurface from opposite directions. In this scenario, the up and down boundaries of the simulation domain are replaced by scattering boundary conditions with specified incident electric field distributions. The relation between the electric field and the input intensity is $E_m = \sqrt{2I_{in,m}Z_0}$ where $Z_0$ is the impedance of free space, and the index $m = 1, 2$ denotes the up and down ports (See Fig. 6(a)), respectively. To calculate the output power ratio $\eta$ in this scenario, the input power is given by $P_{in,m} = a^2 I_{in,m}$, while the output power can be derived by a probe on the output boundary. The probe measures the integration $P_{probe,m} = \iint_C \mathbf{S} \cdot \mathbf{n}$, where $\mathbf{S}$ is the time-averaged Poynting vector, $C$ is the output boundary, and $\mathbf{n}$ is the boundary norm vector. Then, the total output power at port $m$ can be computed by measuring the total power: $P_{out,m} = P_{probe,m} + P_{in,m}$.

## 2. Electric field distribution of the linear bifacial metasurface

When the input intensity is low, the proposed bifacial metasurface works in the linear regime. The linear transmittance spectra of the Fano, Lorentz, and composite metasurface are shown in Fig. 2 in the main paper. The Fano metasurface has a trapped magnetic resonant response, as it is derived by the counter-circulating electric field distributions at the transmission dip, which are shown in the inset of Fig. 2 in the main paper. The electric field distribution on the silicon spheres of the Lorentz metasurface is presented in Fig. S1. The left figure is at the transmission peak $\lambda = 1524$ nm. It can be seen that the direction of the electric field rotates (magnetic response), but now around the z-axis, with a maximum field enhancement of 9.5. All the nanospheres of the Lorentz metasurface have the same electric field distribution and this is a bright magnetic resonant response. On the contrary, the electric field direction of the Lorentz metasurface does not show any characteristic pattern at $\lambda = 1529$ nm, where the Fano metasurface has zero transmittance, and the maximum field enhancement also decreases to 6.5.

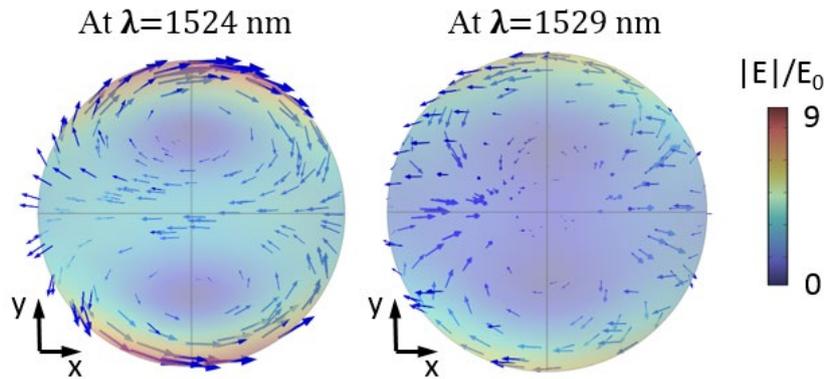

FIG. S1. Direction (blue arrows) and amplitude (color map) of the electric field distribution on the silicon nanoparticles of the linear Lorentz metasurface. The left figure is monitored at the transmission peak $\lambda = 1524$ nm, and the right one is plotted at the zero-transmission wavelength of the Fano metasurface $\lambda = 1529$ nm.

Figure S2 exhibits the electric field direction and enhancement on the Fano metasurface at the transmission peak of the Lorentz metasurface $\lambda = 1524$ nm. As the wavelength deviates from the Fano resonance transmission dip, the maximum field enhancement is reduced to 5, smaller than the $\lambda = 1529$ nm case.

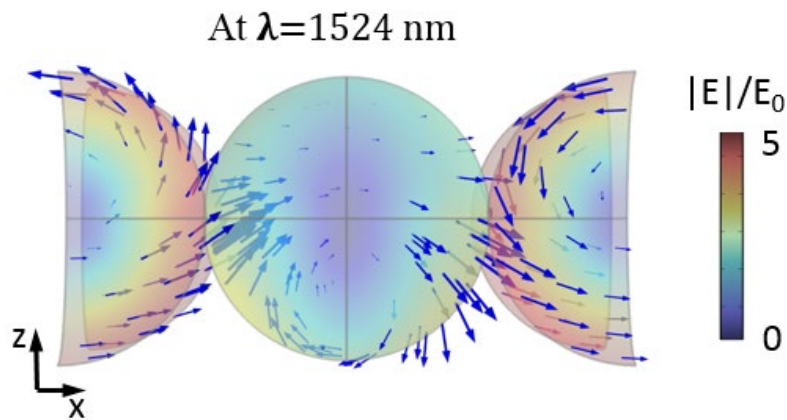

FIG. S2. Direction (blue arrows) and amplitude (color map) of the electric field distribution on the silicon nanoparticles of the linear Fano metasurface at $\lambda = 1524$ nm.

The electric field direction and enhancement on the bifacial metasurface is shown in Fig. S3. The light is fixed at the zero-transmission wavelength $\lambda = 1529$ nm. When the incident light is along the forward direction, it can be seen in Fig. S3(a) that the electric field distribution on the Fano metasurface is almost unchanged, similar to the result presented in the inset of Fig. 2. However, the coupling between the composite layers greatly affects the Lorentz metasurface response. The electric field pattern in Fig. S3(b) is changed compared to the right plot in Fig. S1. Besides, the amplitude is also decreased due to the low transmittance after the incident light propagates through the Fano metasurface. On the other hand, when the incident light is from the backward direction, the electric field amplitude becomes stronger in the Lorentz metasurface (Fig. S3(d)) and comparable to the Fano metasurface (Fig. S3(c)). The distribution pattern of the electric field remains the same in both metasurfaces for backward illumination, while the direction reverses on the Lorentz metasurface.

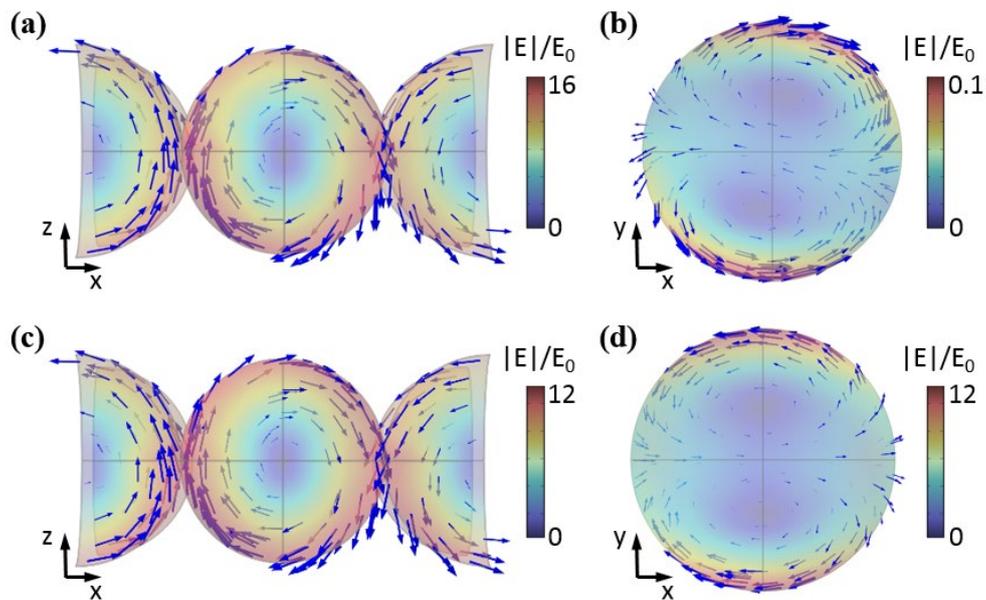

FIG. S3. Direction (blue arrows) and amplitude (color map) of the electric field distribution on the silicon nanoparticles of the linear bifacial metasurface under (a)-(b) forward and (c)-(d) backward incident illumination. The wavelength is fixed to $\lambda = 1529$ nm, where the transmittance is zero. The left field plots ((a)-(c)) are computed on the Fano metasurface, and the right field plots ((b)-(d)) are calculated on the Lorentz metasurface. The color maps are in different scales due to the distinct field amplitude obtained at each resonance.

3. **Electric field distribution in the Lorentz metasurface when nonlinearity is introduced in the proposed bifacial metasurface**

The nonreciprocal transmission in any nonlinear system is mainly due to the asymmetric field distribution. Since the nonlinear permittivity change induced by the Kerr effect is proportional to the local optical intensity, the asymmetric electric field distribution ensures the difference in the nonlinear effective permittivity when the input is injected from opposite sides. The field distribution in the Fano metasurface shows substantial difference in both the maximum field enhancement and the field pattern for opposite illumination directions, as was shown in the inset of Fig. 3 in the main paper. The asymmetry in the field distribution of the Lorentz metasurface under opposite illumination directions is less pronounced. As shown in Fig. S4, the field pattern is almost identical under different illumination directions. The maximum field enhancement is also relative similar, computed to be 11 and 14 under forward and backward incident direction, respectively.

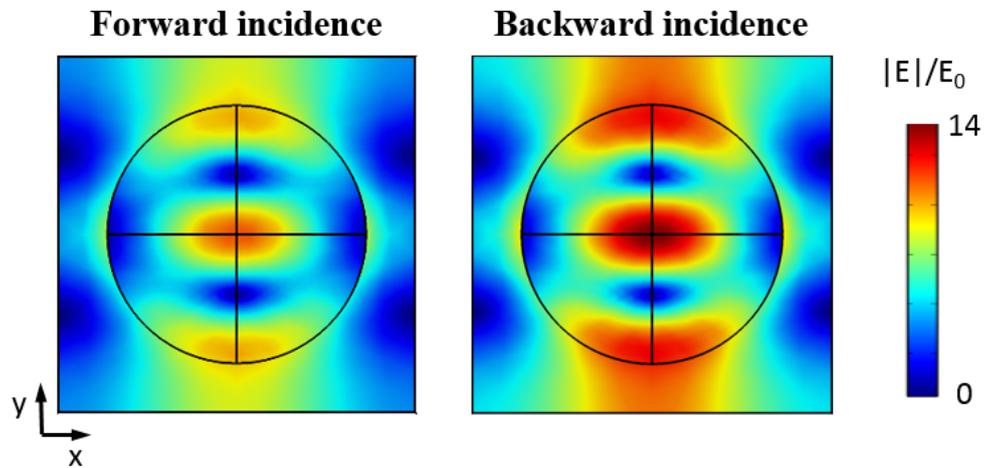

FIG. S4. The electric field enhancement distribution along the x-y plane, across the nanosphere center of the Lorentz metasurface, under a fixed input intensity $I_0 = 2.99$ MW/cm$^2$.